\documentclass[11pt,prd,reprint,onecolumn,notitlepage,nofootinbib]{revtex4-1}

\usepackage{amsmath}
\usepackage{amsfonts}
\usepackage{graphicx}
\usepackage{aas_macros}
\usepackage{color}
\usepackage{listings}

\begin{document}

\title{Low-rank approximations for large stationary covariance matrices, as used
in the Bayesian and generalized-least-squares analysis of pulsar-timing data}
\author{Rutger van Haasteren and Michele Vallisneri}
\address{Jet Propulsion Laboratory, California Institute of Technology, Pasadena, California 91109}
\date{July 11, 2014}

\begin{abstract}
	Many data-analysis problems involve large dense matrices that describe the covariance of stationary noise processes; the computational cost of inverting these matrices, or equivalently of solving linear systems that contain them, is often a practical limit for the analysis.
    We describe two general, practical, and accurate methods to approximate stationary covariance matrices as low-rank matrix products featuring carefully chosen spectral components. These methods can be used to greatly accelerate data-analysis methods in many contexts, such as the Bayesian and generalized-least-squares analysis of pulsar-timing residuals.
\end{abstract}

\maketitle

\section{Introduction}
\label{sec:intro}

For various applications in the mathematical theory of Gaussian processes, in the field of machine learning, and (for these authors especially) in the data analysis of timing residuals from pulsars, one needs to take a matrix inverse of the form $(N + K)^{-1}$, where $N$ is a very large diagonal matrix and $K$ is a very large dense matrix obtained by evaluating a stationary correlation function $k$ at a set of observation times $t_i$:
\begin{equation}
K_{ij} = k(t_i,t_j) = C(t_i - t_j) = C(\tau).
\end{equation}
In the case of pulsar timing, the diagonal of the matrix $N$ would contain the measurement errors for the times of arrival of the pulses, and $C(\tau)$ would describe correlated timing noise with a power-law spectrum $S(f)$, which is present in many pulsar-timing datasets \cite{2010ApJ...725.1607S}; by way of the Wiener--Khinchin theorem,
\begin{equation}
\label{eq:wiener}
C(\tau) = \int_0^\infty\! \mathrm{d}f\, S(f) \cos (2 \pi f \tau).
\end{equation}
In the Bayesian analysis of pulsar-timing \emph{residuals} \cite[the times of arrival minus their theoretical model; see][and references therein]{2009MNRAS.395.1005V,2014arXiv1404.1267A,2014arXiv1407.1838V} the inverse $(N + K)^{-1}$ is required to evaluate the likelihood of the data against a Gaussian-process model of its noise-like components \cite{2014arXiv1407.1838V}; in the generalized--least-squares analysis of residuals \cite[][]{2011MNRAS.418..561C}, the inverse enters the normal equations of the least-squares problem.

Taking the inverse of an $n \times n$ matrix is an $O(n^3)$ operation, which becomes forbidding for large $n$. A popular technique to speed up the inverse is to approximate the dense matrix $K$ as a low-rank product $F \Phi F^T$, where $F$ is $n \times m$ (with $m \ll n$) and $\Phi$ is diagonal (and $m \times m$). The Woodbury matrix identity \cite{doi:10.1137/1031049} then yields
\begin{equation}
(N + K)^{-1} \simeq (N + F \Phi F^T)^{-1} = N^{-1} - N^{-1} F (\Phi^{-1} + F^T N^{-1} F)^{-1} F^T N^{-1},  
\end{equation}
where the dominant cost is the $O(m^3)$ inversion of $\Phi$.
The generalized least-squares approach \cite{2011MNRAS.418..561C} uses a Cholesky decomposition \cite{trefethen1997numerical} of $(N + K)$, which can also be obtained efficiently as a low-rank update \cite{Goldfarb2004, Goldfarb2005, Smola2004}.
The problem then is to obtain an adequate approximation in the form $F \Phi F^T$.
A straightforward solution would be the truncated singular-value decomposition \cite{trefethen1997numerical}, which however is itself an $O(n^3)$ operation. Various other approximations have been discussed in the machine-learning literature \cite{rasmussen2006gaussian}.
This paper discusses the limitations of the popular ``Fourier-sum'' prescription used in pulsar timing (Sec.\ \ref{sec:fourier}), and presents two more accurate alternatives (Secs.\ \ref{sec:lowfreqcov} and \ref{sec:cosine}).

In what follows, we characterize the accuracy of low-rank approximations by examining the difference between exact and approximated covariance matrices for power-law spectra of the form $S(f) = f^{-\gamma} \Theta(f - f_L)$, where $\gamma \in [1,12]$ and where the low-frequency cutoff $f_L$ is required to regularize the DC integral $C(0)$; for concreteness, we set $f_L = 1/(10 \, T)$, where $T$ is the total time span of observations.
We also characterize the accuracy in approximating the \emph{effective} covariance matrix that results after subtracting the best-fit constant, linear, and quadratic terms from the residuals. In pulsar-timing data analysis, this subtraction is performed to account for the intrinsic evolution of pulsar spins; the effective covariance matrix is given by $P K P$, where $P$ is the orthogonal projector onto the null space of the \emph{design matrix} given by constant, linear, and quadratic basis vectors evaluated at the observation times \cite{2013MNRAS.428.1147V}. Along with this manuscript, we provide an implementation of these methods in the Python and C programming languages as ancillary files.

\section{Fourier-sum expansion of covariance} \label{sec:fourier}

In the context of pulsar timing, \citet{2013PhRvD..87j4021L} describe a method where correlated timing noise $y$ is described as an explicit sum of Fourier components.
%
\begin{equation}
y(t) = \sum_{\mu=1}^{2q} w_\mu \phi_\mu(t) = \sum_{k=1}^q a_\mu \cos (2 \pi k \, x) + b_\mu \sin (2\pi k \, x)
\quad \text{with} \quad
x = (t - t_0)/T,
\label{eq:fouriersum}
\end{equation}
where $t_0$ and $T$ are the beginning time and duration of the observation.
The $w_\mu$ are given a normal prior probability, leading to a diagonal matrix $\Phi$ where the same variance $\rho_k$ is shared by the cosine and sine of the same frequency. The basis vectors $\phi_\mu(t_i)$ are only approximately orthogonal, because the $t_i$ are not necessarily sampled regularly. In effect, Eq.\ \eqref{eq:fouriersum} describes a Gaussian process with covariance
\begin{equation}
\begin{aligned}
k(t_1,t_2) &= \sum_{\mu\nu} \phi_\mu(x_1) \Phi_{\mu\nu} \phi_\nu(x_2) \\ &=
\sum_k \rho_k [\cos(2 \pi k \, x_1) \cos(2 \pi k \, x_2) + 
               \sin(2 \pi k \, x_1) \sin(2 \pi k \, x_2)] \\
           &= \sum_k \rho_k \cos[2 \pi k (t_1 - t_2) / T] =
           	   \sum_k \rho_k \cos(2 \pi f_k \tau)
\label{eq:lentati}
\end{aligned}
\end{equation}
where $f_k = k/T$, for $k = 1, \ldots, q$. If we set $\rho_k = S(f_k) \Delta f = S(f_k) / T$, the sum produces a crude but practically useful approximation for the integral of Eq.\ \eqref{eq:wiener}.
Thus we obtain a low-rank approximation $K \simeq F \Phi F^T$ where $F_{i\mu} = \phi_\mu(x_i)$ and $\Phi_{\mu\nu} = \delta_{\mu\nu} \rho_\mu$. Similar approximations have been studied in different contexts in the literature \cite{Paciorek:2007:JSSOBK:v19i02,ChalupkaWM13,Lazaro-GredillaCRF10}.
\begin{figure}
\includegraphics[width=0.8\textwidth]{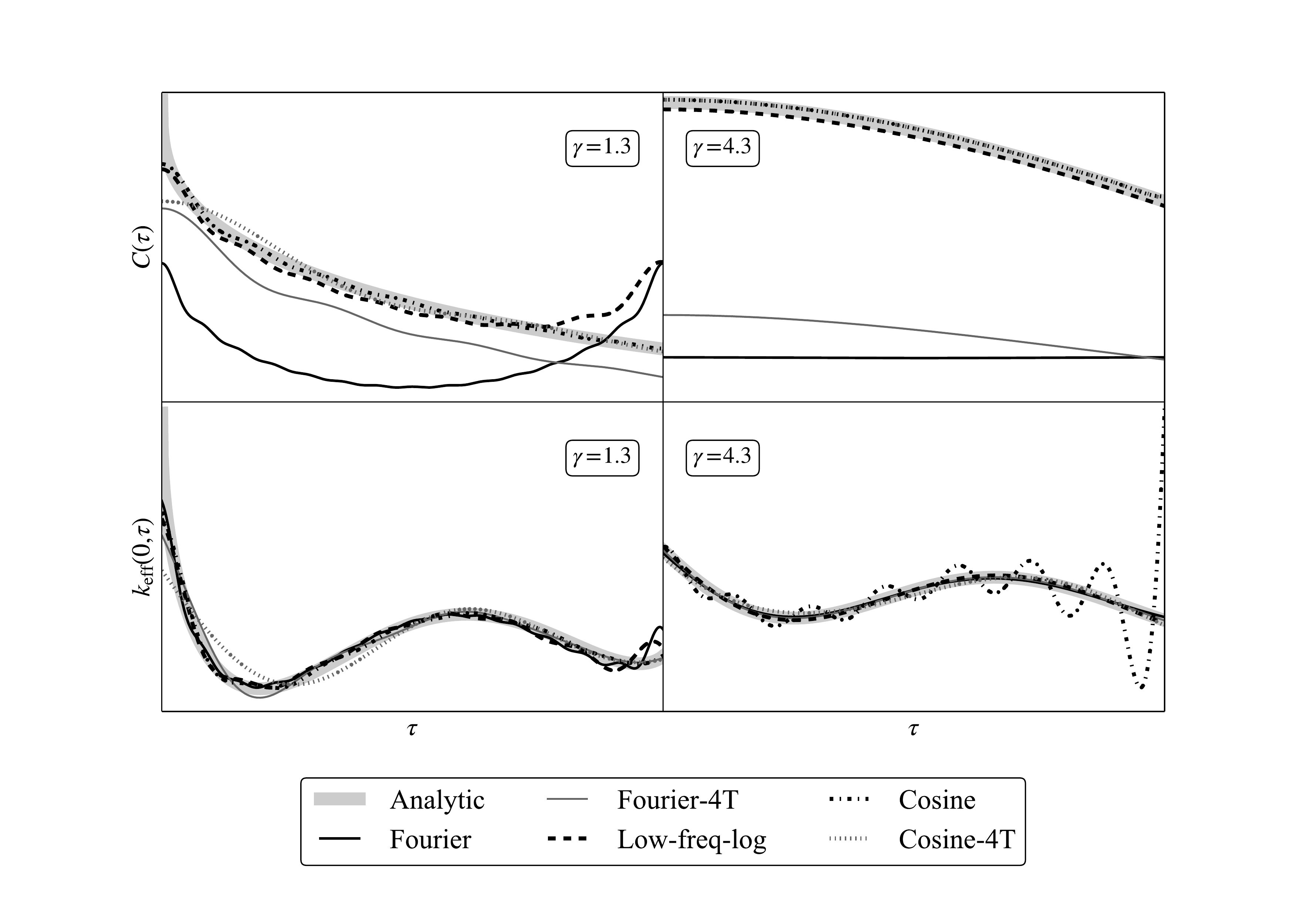}
\caption{
Top: exact (thick gray) and approximated (all others, see legend) covariance function $C(\tau)$ for
$S(f) = f^{-\gamma} \Theta[f - 1/(10 T)]$ with $\gamma = 1.33$ (left) and $\gamma = 4.33$ (right).
Bottom: exact and approximated \emph{post-subtraction} correlation function $k_\mathrm{eft}(0,\tau)$ for the same models.
}
\label{fig:ctau}
\end{figure}

\begin{figure}
\includegraphics[width=0.8\textwidth]{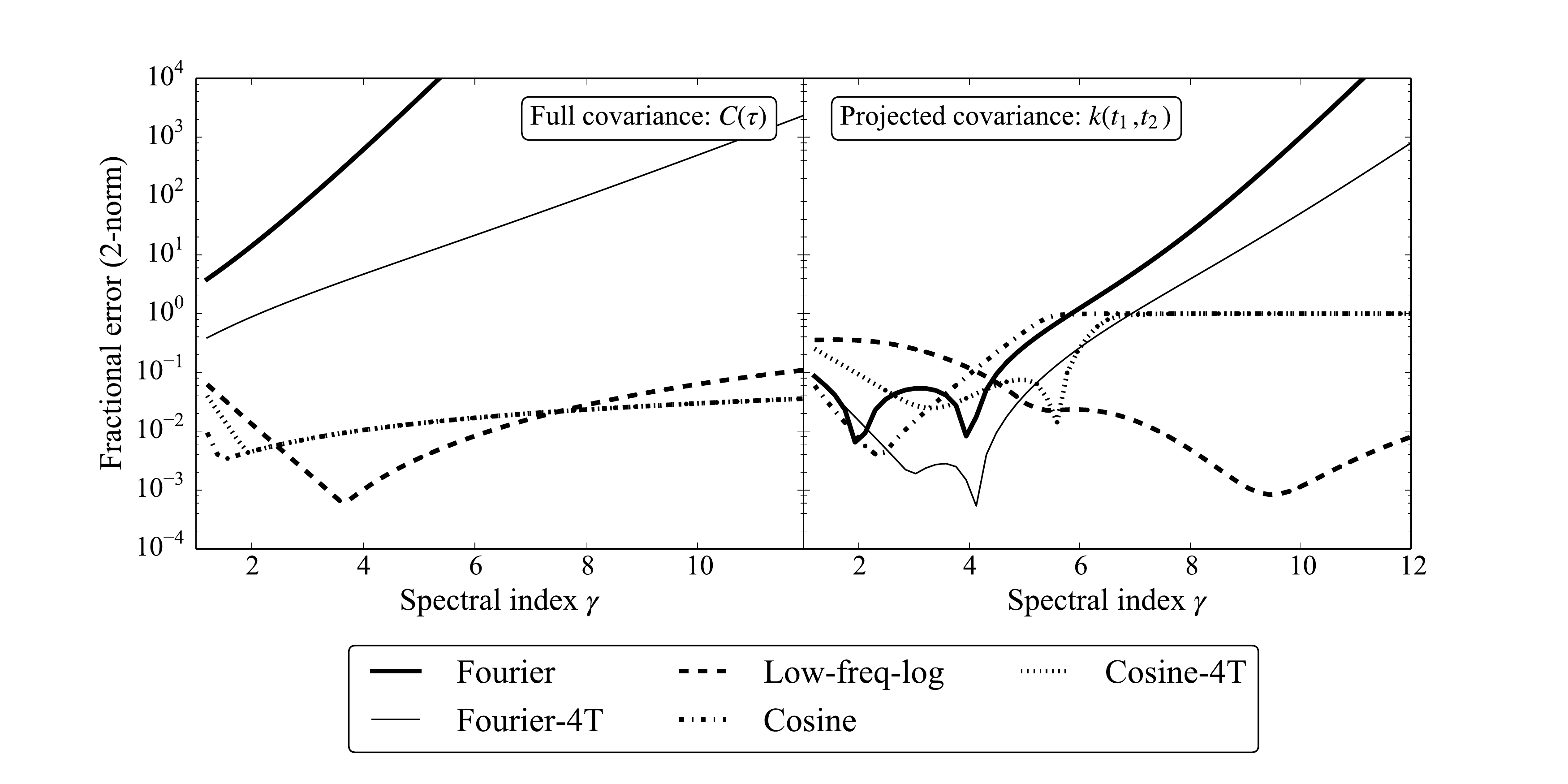}
\caption{
Fractional approximation error as a function of $\gamma$ for all approximation schemes discussed in this paper (see legend), before (left) and after (right) the subtraction of constant, linear, and quadratic trends. The fractional error is defined as the induced Euclidean norm (the 2-norm) of the difference between exact and approximated $K_{ij}$ matrices, divided by the norm of the exact $K_{ij}$; it is evaluated for 1,000 equally spaced $t_i$ between $0$ and the maximum observation time $T$.}
\label{fig:normcomb}
\end{figure}

In Fig.\ \ref{fig:ctau} we show the exact and approximated $C(\tau)$ for $\gamma = 1.33$ and $4.33$ (typical values in pulsar-timing applications), as well as the exact and approximated \emph{post-projection} correlation function $k_{\rm{eff}}(0,\tau)$ [in this second case, we cannot show $C(\tau)$ because the correlation function is not stationary after the subtraction of linear and quadratic trends]. The exact $C(\tau)$ and $k_{\rm{eff}}(0,\tau)$ appear as thick gray curves, while their Fourier-sum approximations are shown as thin dark continuous curves. The thin light continuous curves show the ``$4 T$'' Fourier-sum approximation obtained by taking $f_k = k / (4T)$. The other curves correspond to the novel approximation schemes described in Secs.\ \ref{sec:lowfreqcov} and \ref{sec:cosine}. For all of them we use seven frequency components, corresponding to $F$ and $\Phi$ matrices with 14 columns. We are purposefully including fewer components that would be used in practice, so that the ways in which different approximations fail can be distinguished in the figure.

In Fig.\ \ref{fig:normcomb} we display the fractional approximation error, defined as the induced Euclidean norm (the 2-norm) of the difference between exact and approximated $K_{ij}$ matrices, divided by the norm of the exact $K_{ij}$; the fractional error is shown as a function of $\gamma$ both before (left) and after (right) projection, and it is evaluated for 1,000 equally spaced $t_i$ between $0$ and the maximum observation time $T$. Again the thin dark and light continuous curves show the Fourier-sum approximation and its ``$4 T$'' version. In this case, we use 30 frequency components for all approximation schemes, corresponding to $F$ and $\Phi$ matrices with 60 columns. This many components appear to saturate the accuracy of the schemes, at least for $\gamma$ larger than $3\mbox{--}4$.

The top panels of Fig.\ \ref{fig:ctau} show that the Fourier-sum expansion is very inaccurate in approximating $C(\tau)$, because it lacks a constant offset term. In addition, the Fourier-sum expansion will always produce a \emph{periodic} covariance function, although this is remedied in the ``$4 T$'' version by pushing the periodic boundary way beyond the $[0,T]$ range over which the covariance function would be sampled. As shown in the bottom panels of Fig.\ \ref{fig:ctau}, these handicaps disappear for the post-subtraction effective covariance, especially with the ``$4 T$'' Fourier sum, because the subtraction removes the constant $C(\tau)$ offset and makes the approximation non-periodic. Figure \ref{fig:normcomb} confirms this qualitative picture across the entire range of $\gamma$, and shows that for the projected covariance the ``$4 T$'' Fourier sum is actually rather competitive among the approximations discussed in this paper.

\section{Low-frequency-logarithmic expansion of covariance} \label{sec:lowfreqcov}

\begin{figure}
\includegraphics[width=0.7\textwidth]{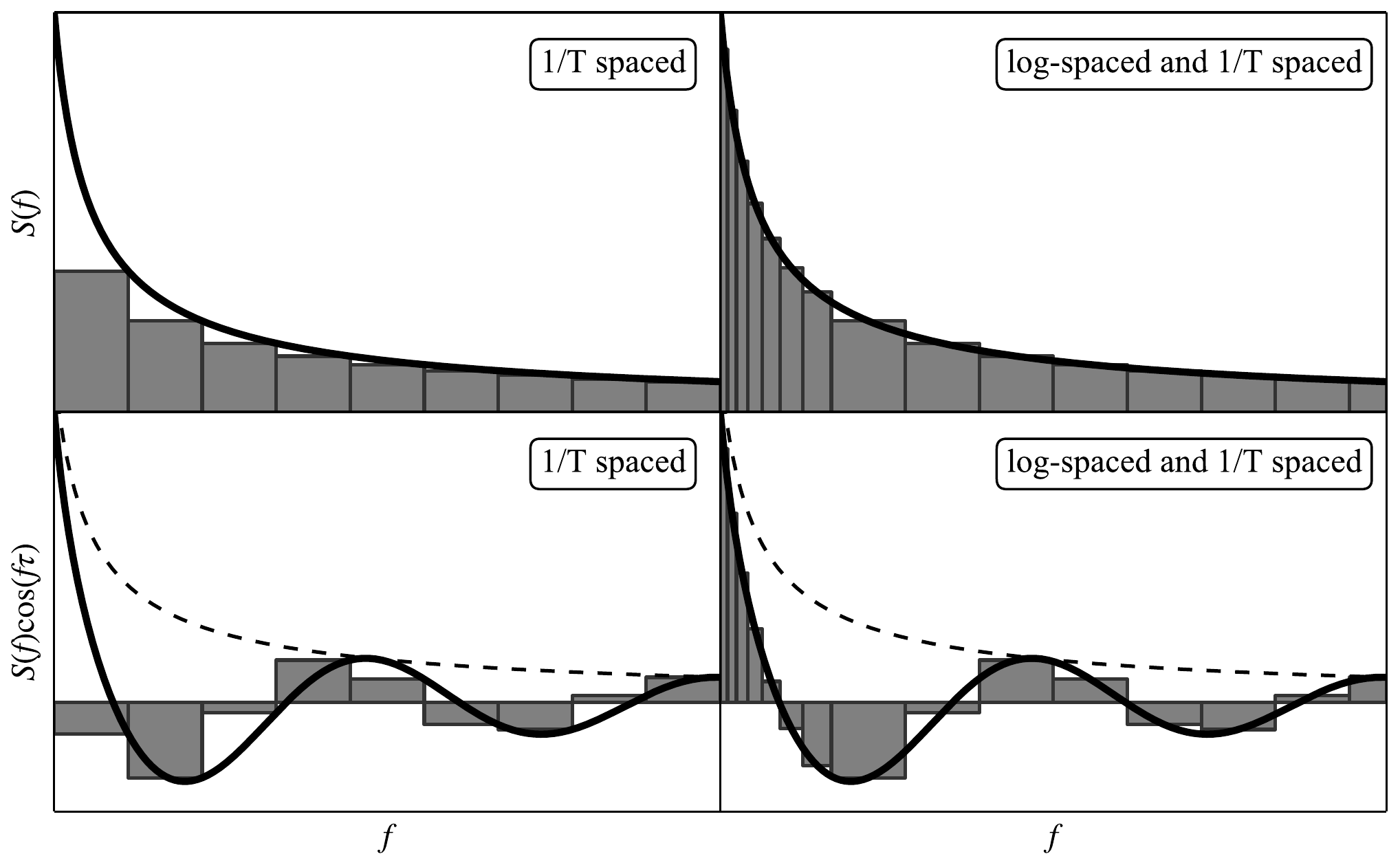}
\caption{Approximating the Wiener--Khinchin integral $C(\tau)$ as a sum of $S(f_i)$ over the frequencies $f_i$, for $\tau = 0$ (top) and for a representative nonzero $\tau$ (bottom). For ``red'' spectra that blow up at low frequencies, a sum with equally spaced bins (left panels) is improved by inserting more bins at low frequencies (right panels).}
\label{fig:freqcov}
\end{figure}

The top-left panel of Fig.\ \ref{fig:freqcov} visualizes the approximation of Eq.\ \eqref{eq:lentati} for the integral $C(0) = \int S(f) \, \mathrm{d}f$ as the rectangular sum $\sum_{k=1}^q S(f_k) \Delta f$, with $f_k = k/T$ and $\Delta f = 1/T$. As in the figure, we may associate the $S(k/T)$ samplings of the spectral density with the right edges of the rectangles; however, we may also center the rectangles, showing that the $[0,1/2T]$ interval -- just where red spectra blow up -- is effectively left uncovered. Our improved approximation (first conceived by the first of these authors, RvH) consists in adding a few logarithmically spaced rectangles to this low-frequency interval, as shown in the right panel of Fig.\ \ref{fig:freqcov}, and in using a slightly more sophisticated integration rule. The bottom panels of Fig.\ \ref{fig:freqcov} show the approximation of the oscillatory $C(\tau)$ integral, for a representative choice of $\tau$, by way of Eq.\ \eqref{eq:lentati} and of our improved scheme.

Specifically, we rewrite the Wiener--Khinchin integral as the sum
\begin{equation}
\label{eq:splitwiener}
C(\tau) = \int_0^\infty\! \mathrm{d}f\, S(f) \cos (2 \pi f \tau) \approx
\int_{\log f_L}^{\log f_M}\! \mathrm{d}(\log f)\,  f S(f) \cos (2 \pi f \tau) +
\int_{f_M}^{f_H}\! \mathrm{d}f\,  S(f) \cos (2 \pi f \tau)
\end{equation}
where the low- and high-frequency cutoffs $f_L$ and $f_H$, as well as the turnover frequency $f_M$, are chosen depending on the specific spectrum $S(f)$ (or family of spectra) under examination. We approximate the two subintegrals in Eq.\ \eqref{eq:splitwiener} with the extended Simpson rule \cite{1992nrca.book.....P}
\begin{equation}
\int_{x_1}^{x_n} g(x) \, \mathrm{d} x \approx \frac{2\Delta x}{6} \bigg(g(x_1)+4g(x_2)+2g(x_3)+4g(x_4)\ldots+4g(x_{n-1})+g(x_n)\bigg) \quad \text{(with odd $n$)},
\label{eq:simpsum}
\end{equation}
where $x$ is $\log f$ and $f$ respectively for the two subintegrals, and where $g$ and the edges and spacings of the $x_i$ are chosen accordingly. Incorporating both the logarithmic bin size and the Simpson-rule coefficients in the weights $w_i$, we can then write
\begin{equation}
    C(\tau) \approx \sum_i w_i S(f_i) \cos(2\pi f_i \tau),
\end{equation}
which summarizes our first improved scheme for approximating the covariance matrix.

As shown in Figs.\ \ref{fig:ctau} and \ref{fig:normcomb}, the low-frequency-logarithmic expansion improves on the Fourier-sum prescription, especially for larger $\gamma$. Just as for Fourier sums, accuracy is better for the post-subtraction covariance matrix than for the unsubtracted $C(\tau)$. To produce these plots, the integration limit $f_L$ of Eq.\ \eqref{eq:splitwiener} was matched with the cutoff frequency $f_L = 1/(10 T)$ of the exact covariance function.

\section{Time-domain cosine expansion of covariance} \label{sec:cosine}

Our second improved scheme is motivated by the observation that
the Wiener--Khinchin integrand is very oscillatory: this is challenging for reduced-rank approximations that represent the integral of Eq.\ \eqref{eq:wiener} as a sum over a small set of evaluations of the integrand, unless $S(f)$ declines so rapidly with $f$ that only a few cycles of $\cos (2 \pi f \tau)$ are relevant.

By contrast, the integrated $C(\tau)$ is usually a smooth function of $\tau$. With this in mind, a useful way to interpret Eq.\ \eqref{eq:lentati} is to view it as a truncated series expansion for the time-domain function $C(\tau)$, rather than a discrete-sum approximation for the Fourier-domain integral \eqref{eq:wiener}. In particular, if we take $f_k = k/2T$, with $k = 0, \ldots, q$, Eq.\ \eqref{eq:lentati} corresponds to the standard discrete cosine transform \cite{1992nrca.book.....P} with coefficients given by
\begin{equation}
\rho_j = \frac{2}{p} \sum_{k=1}^p {\big.}' C\biggl(\frac{(k - 1/2)T}{p}\biggr)
    \cos\biggl(\frac{\pi j(k - 1/2)}{p}\biggr).
\label{eq:coefficients}
\end{equation}
Here $p$ can be equal to $q$, or larger, in which case we are using the first $q + 1$ terms of the $p$-point transform. The prime over the summation symbol indicates that $\rho_0$ must be multiplied by $1/2$. The cosine transform can also be viewed as an interpolation of $C(\tau)$ in terms of Chebyshev polynomials $T_j(\xi)$, where $\xi = \cos (\pi \tau / T)$. This interpretation offers some guarantees about the optimality of the cosine-series approximation \cite{1992nrca.book.....P}, although these would be expressed relative to the transformed time coordinate $\xi$.
%
%
%
%
%

In fact, we do not even need to work with the integrated $C(\tau)$, but we can obtain the $\rho_j$ directly from the power spectral density. In the limit $p \rightarrow \infty$, replacing Eq.\ \eqref{eq:wiener} in Eq.\ \eqref{eq:coefficients} yields
\begin{equation}
\label{eq:cosinecoeff}
\rho_j = (-1)^{j+1} \int S(f) \frac{4 f T \sin(2 \pi f T)}{\pi j^2 - 4 \pi f^2 T^2} \mathrm{d}f,
\end{equation}
which describes the coefficients for our second improved scheme, which was first conceived by the second of these authors (MV).

This scheme and the logarithmic-spacing scheme of Sec.\ \ref{sec:lowfreqcov} are related to the ``sparse spectrum'' method of L\'azaro--Gredilla et al.\ \cite{Lazaro-GredillaCRF10}, in that we seek to optimize the choice of the frequencies in the last row of Eq.\ \eqref{eq:lentati}; however, L\'azaro--Gredilla and colleagues do so by straight numerical optimization, while our recipes are motivated by approximating the covariance integral as a finite sum, and by expanding the covariance, seen as a time-domain function, as a truncated cosine series.

The cosine expansion can be used for spectral estimation with a very general representation of $S(f)$.
If $S(f)$ is provided as an interpolant $\sum_m S(f_m) I_m(f - f_m)$, where $I_m(f - f_m)$ is an interpolation kernel, such as a boxcar or triangle function for nearest-neighbor and linear interpolation respectively, then the coefficients $\rho_j$ are related to the $S(f_m)$ by the linear transformation
\begin{equation}
\rho_j = M_{jm} S(f_m), \quad \text{with} \quad
M_{jm} = (-1)^{j+1} \int I_k(f - f_m) \frac{4 f T \sin(2 \pi f T)}{\pi j^2 - 4 \pi f^2 T^2} \mathrm{d}f,
\end{equation}
where the matrix $M_{jm}$ is constant once the $f_m$ are chosen.

As shown in Figs.\ \ref{fig:ctau} and \ref{fig:normcomb}, the cosine expansion is remarkably accurate for the unsubtracted $C(\tau)$, since it includes a constant term and it is not limited to representing periodic functions over $[0,T]$ (rather, it represents functions with periodic first derivative). However, it suffers from oscillatory error (a form of Gibbs phenomenon \cite{bracewell1978fourier}) that becomes more evident in the post-subtraction covariance. That error is alleviated by replacing $T$ with a small multiple in Eqs.\ \eqref{eq:lentati} and \eqref{eq:cosinecoeff}.

\section{Conclusions} \label{sec:conclusions}

In this paper we investigate the low-rank approximation of stationary covariance matrices (Sec.\ \ref{sec:intro}), with a special focus on applications in pulsar-timing data analysis, and we quantify the accuracy of the commonly used ``Fourier-sum'' approximation \cite{2013PhRvD..87j4021L} (Sec.\ \ref{sec:fourier}) and of two novel methods proposed here (Secs.\ \ref{sec:lowfreqcov} and \ref{sec:cosine}), by way of the matrix norm of the covariance-matrix error. We have included code demonstrations in the Python and C programming languages in the ancillary files of this manuscript.

The Fourier-sum approximation is accurate only when the covariance matrix is projected onto a subspace orthogonal to second-order polynomials in time (which is usually the case in pulsar timing because the linear and quadratic evolution of the pulsar frequency are fit as model parameters), and if the power spectral density represented by the matrix is not too step at low frequencies (for a power law, this corresponds to spectral index $\gamma \lesssim 7$). Even then, the lack of low-frequency components in the Fourier-sum expansion biases the estimation of timing-model parameters such as quadratic spindown; conversely, it requires the artificial addition of second-order polynomials to the timing model when these are not naturally present, as in the case of dispersion-measure variations \cite{2014MNRAS.437.3004L}.

We resolve these maladies in two novel schemes. In the first, we generalize the Fourier-sum expansion by interpreting it as the discrete approximation of the covariance integral, which prompts the addition of low-frequency terms with logarithmic spacing. Only a few terms are needed to improve low-frequency coverage. In the second, we repurpose and modify the Fourier-sum expansion to encode the cosine expansion of the time-domain covariance function (or equivalently, as its Chebyshev interpolation in a transformed time coordinate).

Both schemes vastly outperform the Fourier-sum method in approximating the unprojected covariance, as well as the projected covariance for steep red spectra. 
Which of the two new methods performs best and how many frequency components are needed depends on the specifics of the random process that is being approximated, and should be determined case by case.
Compared to the performing matrix algebra on full-sized covariance matrix, our low-rank approximations greatly reduce memory requirements and computational costs, often by orders of magnitude.

\paragraph*{Acknowledgments.}
    RvH was supported by NASA Einstein Fellowship grant
    PF3-140116. MV was supported by the Jet Propulsion Laboratory RTD program.
    This work was carried out at the Jet Propulsion Laboratory, California
    Institute of Technology, under contract to the National Aeronautics and
    Space Administration. Copyright 2014 California Institute of Technology.
    Government sponsorship acknowledged.

\bibliography{cheby}

\end{document}